\documentclass[12pt]{article}
\usepackage{amsfonts}
\usepackage{amssymb}
\def\bsh{\backslash}

\def\adt{\dot \alpha}

\newfont{\bbbold}{msbm10} %{scaled \magstep1}
\def\bbA{\mbox{\bbbold A}}

\def\bbC{\mbox{\bbbold C}}

\def\bbF{\mbox{\bbbold F}}
\def\bbG{\mbox{\bbbold G}}

\def\bbZ{\mbox{\bbbold Z}}
\def\cA{\cal A}

\def\cO{{\cal O}}

\newfont{\goth}{eufm10 scaled \magstep1}

\def\gl{\mbox{\goth l}}

\def\gp{\mbox{\goth p}}

\def\gs{\mbox{\goth s}}

\def\a{\alpha}

\def\d{\delta}\def\D{\Delta}
\def\e{\epsilon}

\def\l{\lambda}\def\L{\Lambda}
\def\m{\mu}

\def\p{\pi}
\def\P{\Pi}

\def\t{\tau}

\def\be{\begin{equation}}\def\ee{\end{equation}}
\def\bea{\begin{eqnarray}}\def\eea{\end{eqnarray}}
\def\ba{\begin{array}}\def\ea{\end{array}}

\def\del{\partial}

\def\str{\rm str}
\def\xz{\times}

\def\nab{\nabla}

\def\del{\partial}
\def\sdet{{\rm sdet}}
\def\str{{\rm str}}
\def\pheq{&\phantom{=}&}
\let\la=\label

\def\nn{\nonumber}
\def\bd{\begin{document}}
\def\ed{\end{document}}
\def\bea{\begin{eqnarray}}
\def\eea{\end{eqnarray}}
\def\ba{\begin{array}}
\def\ea{\end{array}}
\def\ft#1#2{{\textstyle{{\scriptstyle #1}\over {\scriptstyle #2}}}}
\def\fft#1#2{{#1 \over #2}}
\newcommand{\eq}[1]{(\ref{#1})}
\def\eqs#1#2{(\ref{#1}-\ref{#2})}
\def\det{{\rm det\,}}
\def\tr{{\rm tr}}
\newcommand{\ho}[1]{$\, ^{#1}$}
\newcommand{\hoch}[1]{$\, ^{#1}$}
\newcommand{\tamphys}{\it\small Center for Theoretical Physics,
Texas A\&M University, College Station, TX 77843, USA}
\newcommand{\newton}{\it\small Isaac Newton Institute for Mathematical
Sciences, Cambridge, UK}
\newcommand{\kings}{\it\small Department of Mathematics, King's College,
London, UK}
\newcommand{\lapp}{\it\small LAPP, Annecy, France}
\newcommand{\HA}{{\Bbb H}\hskip-4.1pt {\Bbb A}}
\newcommand{\HR}{{\Bbb H}\hskip-4.7pt {\Bbb R}}
\newcommand{\dslash}{\partial\hskip-6.28pt /}
\begin{document}
 \thispagestyle{empty}

 \hfill{KCL-MTH-01-19}

  \hfill{\today}
\vspace{20pt}

 \begin{center}
{\Large{\bf AdS/SCFT in Superspace}} \vspace{30pt}\\ {\large P.S. Howe and P.C. West}
\\ \vspace{15pt} Department of Mathematics

King's College London UK

\vspace{60pt}
 {\bf Abstract}
\end{center}
A discussion of the AdS/CFT correspondence in IIB is given in a
superspace context. The main emphasis is on the properties of SCFT
correlators on the boundary which are studied using harmonic
superspace techniques. These techniques provide the easiest way of implementing the superconformal Ward identities. The Ward identities, together with analyticity, can be used to give a compelling argument in support of the non-renormalisation theorems for two- and three-point functions, and to establish the triviality of extremal and next-to-extremal correlation functions. Further simplifications in other correlators are also implied. The OPE is also briefly discussed.

{\vfill\leftline{}\vfill
\vskip	10pt
\footnoterule
\footnotesize{Talk given by P.S. Howe at the G\"{u}rsey Memorial Conference II {\sl M-theory and Dualities}, Bogazici University, Istanbul, June 19-23, 2000.}
\pagebreak
\setcounter{page}{1}

%%%%%%%%%%%%%%%%%%%%%%%%%%%%%%%%%%%%%%%%%%%%%%%%%%%%%%%%%
\section{Introduction}

One of the key points of the AdS/CFT conjecture \cite{maldacena} is that the
isometry group of AdS spacetime coincides with the conformal group
of the boundary Minkowski space. In the supersymmetric context,
these groups are extended to supergroups, and their actions as
isometries and superconformal transformations are most easily
presented in superspace. In this paper we shall be mainly concerned
with IIB supergravity on $AdS_5\xz S^5$ and its relation to $N=4$
super Yang-Mills theory (SYM) on the boundary. In this case the
supergroup in question is $PSU(2,2|4)$.

A brief discussion of the basic geometrical set-up and the KK
supermultiplets is given in section 2 and the rest of the paper is devoted to a
review of the analytic superspace approach to correlation functions
of the composite operators which correspond to the KK
supermultipets in the AdS/CFT context.

%%%%%%%%%%%%%%%%%%%%%KK SUPERMULTIPLETS%%%%%%%%%%%%%%%%%%%%%%%%%%%%%%%

\section{KK supermultiplets}

The basic geometrical set-up can be summarised by the following
diagram

\be
\ba{ccc} AdS^{5,5|32} & \rightsquigarrow & M^{4|16}\xz S^5 \\ &&\\
\Big\downarrow && \Big\downarrow \\ &&\\ AdS^{5|32}
&\rightsquigarrow & M^{4|16} \ea \ee

where the squiggly arrows denote passing to the boundary. Each of
these superspaces is a coset space of the supergroup $PSU(2,2|4)$,
with the notation indicating the (even$|$odd) dimensions \cite{adsspace, hesh}. Thus
$AdS^{5,5|32}$ denotes the superspace whose body is $AdS_5\xz S^5$
and which has 32 odd dimensions while $AdS^{5|32}$ is the $D=5,
N=8$ superspace which has $AdS_5$ for its body and which has  32
odd dimensions. The boundary space on the bottom row is $D=4, N=4$
Minkowski superspace. The isotropy group of $AdS^{5,5|32}$ is
$Spin(1,4)\xz USp(4)$ while that of $AdS^{5|32}$ is $Spin(1,4)\xz
SU(4)$ and the former fibres over the latter with fibre $S^5$. The
boundary superspaces are most easily described by viewing
$PSU(2,2|4)$ as the $D=4, N=4$ superconformal group. The generators
of the corresponding Lie superalgebra are $\{D,P,K,M,N;Q,S\}$
standing for dilations ($D$), translations ($P$), Lorentz
transformations ($M$), internal $SU(4)$ transformations ($N$), and
$Q$ and $S$-supersymmetry transformations. The isotropy group of
$M^{4|16}$ is generated by $\{D,K,M,N;S\}$ while the isotropy group
of $M^{4|16}\xz S^5$ is generated by $\{D,K,M,N';S\}$, where $N'$
denotes the generators of the $USp(4)$ subgroup of $SU(4)$.

Linearised type IIB supergravity is described by a chiral field
strength superfield $\bbA$ satisfying an additional fourth-order
constraint which is schematically of the form $D^4\bbA\sim \bar
{D}^4\bar{\bbA}$ \cite{hw1}. This is the case for both  Minkowski and
$AdS^{5,5|32}$ backgrounds \cite{hesh}. In the latter case we can expand this
superfield in harmonics on $S^5$ to obtain a family of chiral
superfields on $AdS^{5|32}$ which fall into symmetric traceless
representations of $SO(6)$, the isometry group of $S^5$. Passing to
the boundary, $D=4,N=4$ super Minkowski space, we find a family of
$D=4, N=4$ chiral field strength superfields which are now
off-shell, and which also satisfy fourth-order constraints. These
fields can be expressed in terms of a family of unconstrained
Grassmann-analytic prepotentials on an appropriate harmonic
superspace, which will be discussed below, and these prepotentials
couple naturally to composite operators of the boundary SCFT which
are both G-analytic and harmonic analytic \cite{hesh}. In fact they are
operators of the form $\tr(W^p),\ p=2,3,\ldots$ where $W$ is the
SYM field strength superfield. This identification of the KK multiplets \cite{gm} and SCFT multiplets was first made in \cite{af}.

%%%%%%%%%%%%%%%%%%%%%%%%HARMONIC SUPERSPACE%%%%%%%%%%%%%%%%%%%%%%%%%
\section{Harmonic superspace}

In extended supersymmetry  flat harmonic superspaces
\cite{gikos1} are superspaces of the form $M_H=M\xz \bbF$ where
$M$ is the corresponding Minkowski superspace and $\bbF$ is a coset
space of the internal symmetry group which is chosen to be a
particular type of compact, complex manifold. For example, in
$D=4$, the internal symmetry group is $SU(N)$ and the possible
internal manifolds of interest are the flag manifolds
$\bbF_{k_1,\ldots k_{\ell}}=S(U(k_1)\xz U(k_2-k_1)\ldots
U(N-k_{\ell}))\bsh SU(N)$, where $k_1<k_2\ldots <k_{\ell}<N$ are
positive integers \cite{ikno}. In particular the isotropy groups $S(U(p)\xz
U(N-(p+q))\xz U(q))$ define $(N,p,q)$ harmonic superspaces \cite{hh}, a
family of superspaces which generalises in a natural way the
harmonic superspaces for $N=2$ and $N=3$ supersymmetry introduced
by GIKOS, which are respectively $(2,1,1)$ \cite{gikos1} and
$(3,1,1)$ \cite{gikos2} harmonic superspace in the above
notation.\footnote{(3,2,1) superspace was first discussed in
\cite{gikos3}}

The harmonic superspace approach of GIKOS emphasises the group
theoretic aspects of fields defined on such spaces rather than the
holomorphic aspects. For this reason it has become standard
practice to work on the space $\hat M_H=M\xz SU(N)$. This is
equivalent to working on $M_H$ provided that the fields are
restricted to be equivariant with respect to the isotropy group,
$H$,  of the relevant coset space. Such a field is a map $F:\hat
M_H\rightarrow V$, where $V$ is a representation space for the
group $H$, such that $F(z,hu)=M(h)F(z,u)$, for all $u\in SU(N),\
h\in H$ and $z\in M$, where $M(h)$ denotes the action of $H$ on
$V$. An equivariant field of this type defines in a natural way a
section of a vector bundle $E$ over $M_H$ with typical fibre $V$
and the two types of object are in one-to one correspondence.

In this paper we shall be concerned with $D=4$ $(N,p,q)$ harmonic
superspaces for the cases $(4,2,2)$ and $(2,1,1)$. A group element
$u$ is written in index notation as $u_I{}^i$ where the $i$ index
is acted on by $SU(N)$ and the $I$ index by $H$, and the inverse
$u^{-1}$ is denoted $u_i{}^I$. The index $I$ splits under $H$ as
$(r,r')$ where $r=1,\ldots p,\ r'=p+1,\ldots N$, $p$ being either
$2$, for $N=4$, or $1$ for $N=2$. With the aid of $u$ and its
inverse  $SU(N)$ indices $i,j,\ldots$ can be converted into $H$
indices $I,J,\dots$ in an obvious manner. In particular we can
construct the Grassmann odd derivatives $D_{\a I}=(D_{\a r},D_{\a
r'})$ and $\bar D_{\adt}^I=(\bar D_{\adt}{}^r,\bar D_{\adt}^{r'})$.

The derivatives $(D_{\a r},\bar D_{\adt}^{r'})$ anticommute and
so allow the introduction of generalised chiral fields, or
Grassmann-analytic (G-analytic) fields, $F$, which satisfy

\be
D_{\a r}F=\bar D_{\adt}^{r'}F=0 \ee

For any fixed values of the $u's$ the derivatives $(D_{\a
r},D_{\adt}^{r'})$ define a CR structure on $M$; that is, they are
basis vector fields for a subspace of the complexified tangent
bundle $T_c$ such that they anticommute and their complex
conjugates are linearly independent of them at any point in
$M$.\footnote{The r\^{o}le of CR structures in harmonic superspace
was first emphasised by \cite{rs}.} The space $\bbF$ can thus be
viewed as the space of all CR structures of this type on $M$. In
addition, the derivatives can be combined with a subset of the
right-invariant vector fields $D_I{}^J$ on $SU(N)$ to define a CR
structure on $M_H$. The right-invariant vector fields decompose
under $H$ into coset derivatives $D_r{}^{s'}$ (essentially
$\bar\del$ on $\bbF$) and its conjugate $D_{r'}{}^s$, and isotropy
group derivatives $D_r{}^s, D_{r'}{}^{s'}$ ($SU(2)\xz SU(2)$ for
the case $N=4$) and $D_o$ ($U(1)$ in both cases).

The CR structure on $M_H$ is then determined by the derivatives
$(D_{\a r}, D_{\adt}^{r'}, D_r{}^{s'})$. Since the $D_I{}^J$ are
characterised by

\bea D_I{}^Ju_K{}^k&=&\d_K{}^J u_I{}^k-\frac 1N\d_I{}^J u_K{}^k\nn
\\ D_I{}^Ju_k{}^K&=&-\d_I{}^K u_k{}^J + \frac 1N \d_I{}J u_k{}^K
\eea

it is straightforward to verify that this set of derivatives does
indeed specify a CR structure. On $M_H$ we therefore have two types
of analyticity - Grassmann analyticity (G-analyticity) and harmonic
analyticity (H-analyticity), the latter meaning analyticity in the usual sense
on $\bbF$. A field which is both G-analytic and H-analytic will be
called CR-analytic, or simply analytic, for short.

$N=2$ harmonic superspace was used by GIKOS to give an
off-shell version of the hypermultiplet (which cannot be formulated
in ordinary superspace with a finite number of auxiliary fields).
GIKOS also reformulated $N=2$ Yang-Mills theory in harmonic
superspace \cite{gikos1}. In this case there is still a finite number of
auxiliary fields (a triplet of dimension two scalars) but the gauge
group is extended and it is this extension that facilitates the
solution of the $N=2$ SYM superspace constraints in terms of a
dimension zero G-analytic prepotential. The $N=3$ theory can also
be given an off-shell formulation in harmonic superspace but in
this case there is both an infinite number of auxiliary fields and
an extended gauge group \cite{gikos2}. We note that, although the on-shell $N=3$
and $N=4$ SYM theories are the same, there is no known off-shell
version which has manifest $N=4$ supersymmetry.

In the present context we are interested in correlation functions
of gauge-invariant composite operators which are constructed as
polynomials in the field strength superfields. We shall make the assumption that, provided that we
restrict our attention to separated points, it should be possible to
use the equations of motion for the SYM fields and thus we shall
only need to have an on-shell formulation of the underlying theory.
We shall consider both  $N=2$ and $N=4$ SYM theories, partially
because it is sometimes convenient to write the latter in terms of
$N=2$ superfields (in particular, this allows checks to be carried
out in perturbation theory which is not possible using on-shell
$N=4$ superfields), and partially because some of the results we
shall obtain are valid for more general $N=2$ superconformal
theories.

The fields for $N=2$ Yang-Mills are the chiral field strength $W$,
$\bar \nab_{\adt}^i W=0$ and the hypermultiplet $q$, which is best
viewed as a field on $N=2$ harmonic superspace, $M_H=M\xz \bbC
P^1$. Off-shell the latter field is G-analytic, but on-shell it
becomes H-analytic as well, and thus analytic in the above
nomenclature. In the free theory $q=u_1{}^i q_i$, where
$u_I{}^i=(u_1{}^i, u_2{}^i)$ for $N=2$. It is not real, but under a
suitable conjugation transforms into $\tilde q= u_1{}^i \bar q_i$.
In the non-Abelian theory it is still a short multiplet of the
above type but is covariantly G-analytic. However, gauge-invariant
products of the $q$'s and $\tilde q$'s are ordinarily analytic. We
shall denote operators with $p$ powers of $q$ or $\tilde q$ by
$\cO_p$.

For $N=4$ Yang-Mills theory the field strength superfield
$W_{ij}=-W_{ij}$ in super Minkowski space transforms according to
the six-dimensional representation of the internal symmetry group
$SU(4)$, so  $W^{*ij}={1\over2}\e^{ijkl} W_{kl}$, and satisfies $\nab_{\a i} W_{jk} =\nab_{\a [i} W_{jk]}$. 
The claim is that $W_{ij}$ is equivalent to a charge 1 field $W$ on
$M_H$ which is covariantly $G$-analytic and ordinarily
$\bbF$-analytic; it is also real with respect to the real structure
discussed above, where covariantly $G$-analytic means that

\be
\nab_{\a r} W=\nab_{\adt}^{r'} W=0 \ee

with $\nab_{\a r}=u_r{}^i\nab_{\a i}$, etc. This claim is easily
verified \cite{hh}.\footnote{The field strength can also be viewed
as an analytic field on $M\xz U(1)^3\bsh SU(4)$ \cite{b}, but this
description is not so convenient from the point of view of
superconformal transformations if one wishes to take maximal
advantage of G-analyticity.} The convention for $U(1)$ charges is

\be
D_o u_r{}^i= {1\over2}u_r{}^i;\ D_o u_{r'}{}^i=
-{1\over2}u_{r'}{}^i \ee

The gauge-invariant operators $\cO_p:=\tr (W^p)$ are $G$-analytic
in the usual sense and hence analytic. These are the conformal
fields of interest in the Maldacena conjecture. Since they are in
short representations of $SU(2,2|4)$, the integer $p$ cannot be
affected by quantum corrections and so, since this integer
determines the dimension, they do not have anomalous dimensions.
This family of operators was introduced in \cite{hw2} and it has
been shown that it is in one-to-one correspondence with the
Kaluza-Klein spectrum of IIB supergravity on $AdS_5\xz S^5$
\cite{af}. In particular, the family of operators includes the
energy-momentum tensor $T=\cO_2$, first presented as a
harmonic superfield in \cite{hh}. We shall also be able to study
correlators of gauge-invariant products of hypermultiplets in $N=2$
theories; again, these operators are analytic fields.

\section{Correlators in analytic superspace}

Analytic superspaces are related to harmonic superspaces rather as
chiral superspaces are related to Minkowski superspaces. Such
spaces are intrinsically complex and are not coset spaces of the
real superconformal group. From the point of view of manifest
superconformal invariance it is therefore convenient to complexify
both spacetime and the superconformal group (which becomes
$SL(4|N)$). An analytic superfield on harmonic superspace can then
be rewritten as an unconstrained, holomorphic field on analytic
superspace. We emphasise that this is simply a matter of
convenience; we can return to real spacetime by imposing
suitable reality conditions on the coordinates. It is to be noted
that the internal flag space remains unchanged in this
construction. The key point  is that correlation functions of analytic operators must be
analytic in the internal coordinates $y$, because we can write each
operator explicitly as a polynomial in $y$ with coefficients which
depend on the coordinates of Minkowski superspace. On the other
hand, as we shall see below, any analytic invariant has singularities in the $y$'s. This
circumstance places constraints on the correlators which are strong enough in some cases to determine them completely, as conjectured in \cite{hw2}.

We shall be concerned with the analytic superspaces associated with
$(4,2,2)$ and $(2,1,1)$ harmonic superspaces. On these analytic
superspaces the coordinates are

\be
X^{AA'}=\left(\ba{cc} x^{\a\adt}& \l^{\a a'}\\
\p^{a\adt}&y^{aa'}\ea\right) \ee

Here $\a$ and $\adt$ are two-component spinor indices, while $a$
and $a'$ are internal ``spinor'' indices; in $N=2$ these indices
take on one value each and so can be omitted, while in $N=4$ they
take on two values and behave in a similar manner to the spacetime
indices.

An analytic field $\cO_p$ of charge $p$ transforms under
superconformal transformations according to the rule

\be
\d \cO_p=V \cO_p + p\D \cO_p \ee

where the vector field $V=\d X{\del\over\del X}$ is determined by
an infinitesimal superconformal transformation  $\d X$

\be
\d X=B + AX + XD + XCX \ee

and where the function $\D=\str (A+XC)$.

The supermatrix parameters $A,B,C$ and $D$ together make up an
element $\cA$ of the Lie superalgebra $\gs\gl(4|N)$,

\be
\cA=\left(\ba{cc} -A&B\\ -C&D \ea\right) \ee

In the $N=4$ case the term in $\cA$ proportional to the unit matrix
acts trivially on $X$ so that we indeed have an action of
$\gp\gs\gl(4|4)$.

We shall be interested in correlation functions of the form

\be
<p_1\ldots p_n>:=<\cO_{p_1}(X_1)\ldots \cO_{p_n}(X_n)> \ee

The Ward identity for such a correlation function, which states
that it is invariant under superconformal transformations, is as
follows \cite{hw2}:

\be
\sum_{i=1}^n \left(V_i + p_i \D_i\right) <p_1\ldots p_n>=0 \la{wi}
\ee

where the index $i$ refers to the point $X_i$ of the operator
$\cO_{p_i}$.

The basic building block which one uses to analyse such correlators
is the two-point function for charge one operators in the free
theory, that is, $<WW>$ in $N=4$, and $<q\tilde q>$ in $N=2$, where
$q$ is a hypermultiplet and $\tilde q$ its harmonic conjugate. This
two-point function, which we shall refer to as a propagator and denote by $g_{12}$, has the
following form

\be
g_{12}=\cases{{\hat y_{12}\over x_{12}^2},\ {\rm for}\ N=2\cr 
\ \ \ \ \ \ \ \ \ \ \  \cr
{\hat
y_{12}^2\over x_{12}^2},\ {\rm for}\ N=4} \la{prop} \ee

where

\be
\hat y_{12}=y_{12}-\p_{12}x_{12}^{-1}\l_{12} \ee

with $X_{12}:=X_1-X_2$ as usual. This formula is to be understood
in the sense of matrix multiplication with $x^{-1}$ being assigned
subscript indices $\adt\a$.

An arbitrary correlation function of analytic fields, as long as it
is non-vanishing at lowest order in an expansion in the odd
coordinates, can then be expressed as a product of propagators
times a function $F$ of the $n$ coordinates which is
superconformally invariant,

\be
<p_1 p_2\ldots p_n> =\prod_{i<j} (g_{ij})^{p_{ij}} F \ee

The powers $p_{ij}$ must be chosen such that the prefactor
multiplying $F$ on the right-hand side solves the Ward identities
for the given charges; in general there will be different  ways of
doing this.

As we remarked above, any such correlation function must be
analytic in the internal coordinates $y$, because we can write each
operator explicitly as a polynomial in $y$ with coefficients which
depend on the coordinates of Minkowski superspace. On the other
hand, any invariant has singularities in the $y$'s. This
circumstance places constraints on the functional form of $F$ - the
only singularities in $y$ it can have must be those which can be
cancelled by the zeroes in the propagators. It was conjectured in \cite{hw2} that these constraints could be strong enough to completely determine some particular types of correlation functions and this is indeed the case as we shall demonstrate below.

%%%%%%%%%%%%%%%%%%%INVARIANTS%%%%%%%%%%%%%%%%%%%%%%%%%%%%%%%

\section{Invariants}

In \cite{hw3} a method for constructing non-nilpotent invariants in analytic superspace was given. However, it was
pointed out in \cite{ken1} that the $N=4$ invariants given there are in
fact invariant under an additional symmetry, $U(1)_Y$, and therefore under the group
$PGL(4|4)$ and not just $PSL(4|4)$. In \cite{ehw} it was shown that there are non-nilpotent invariants in $N=4$ which are not invariant under $U_Y(1)$. Here we give a brief
account of a systematic method which can be used to construct all the invariants.
It is, in fact, a valid method for all Grassmannians of the form
$\bbG_n(2n)$ where $n$ can be either an integer (bosonic case) or a
pair of integers (super case). It is based on the approach of
\cite{o}

Consider the transformation

\be
\d X_i=B + A X_i + X_i D + X_i C X_i,\ i=1,\ldots n,\ {\rm no\ sum\
on\ i} \ee

where $X_i$ is the supercoordinate $X_i^{AA'}$ of the $i$th point.
We are looking for functions $F(X_1,\ldots X_n)$ which are
invariant under the above. We first solve for translations $B$. If
we change coordinates to $(X_1,X_{1i}),i=2\ldots n$ we find that
$F$ is independent of $X_1$. Now consider the transformation of
$X_{1i}$ under $C$,

\bea
 \d_C X_{1i}&=&X_1C X_1-X_i C X_i \nn\\
 &\phantom{=}&-X_{1i}CX_{1i}+X_1CX_{1i}+ X_{1i} C X_1
 \eea

For the inverse we therefore have

\be
\d X_{1i}^{-1}=C -(C X_1)X_{1i}^{-1}- X_{1i}^{-1}(X_1 C) \ee

At this stage we can regard $F$ as being a function of the $n-1$
inverses $X_{1i}^{-1}$ and change variables to $X_{12}^{-1}$ and
$n-2$ variables $Y_i$ defined by

\be
Y_i:= X_{12}^{-1}-X_{1i}^{-1},\qquad i=1,\ldots n-2 \ee

We have

\be
\d_C Y_i=-(C X_1)Y_i - Y_i(C X_1) \ee

and so the invariance of $F$ under $C$ implies

\bea \d_C F&=&\left(C-(CX_1)X_{12}^{-1}-X_{12}^{-1}(X_1 C)\right)
{\del F\over\del X_{12}^{-1}} \nn \\
 \pheq + \sum_{i=3}^{n}\left(-(C X_1)Y_i - Y_i(C X_1)\right){\del F\over \del
 Y_i}\ \ =\ \ 0
 \eea

Now $F$ is independent of $X_1$ so the above should be valid for
arbitrary values of this coordinate. Taking $X_1=0$ we see that $F$
does not depend on $X_{12}^{-1}$. Thus $F$ depends only on the
$(n-2)$ $Y_i$'s and the residual transformation reduces to linear
transformations of type $A$ and $D$. Hence, if $F$, as a function
of the $Y_i$,  is invariant under the linear $A$ and $D$ symmetries
it will automatically be completely invariant. Now each of the
parameter matrices $A$ and $D$ contain as many odd entries as there
are odd coordinates. Solving the constraints on $F$ due to the odd linear symmetries allows us to
eliminate a further two sets of odd variables and thereby arrive at
a set of $(n-2)$ sets of even coordinates ($x$'s and $y$'s) and
$(n-4)$ sets of odd coordinates ($\l$'s and $\p$'s). The final step
is to construct functions from these which are invariant under
standard bosonic linear symmetries which simply requires the
indices to be hooked up in the right way and the weights to cancel.

From this analysis we can extract a few simple but important
results \cite{ehw}. There are no invariants for $N=2$ and $N=4$ analytic superspaces at two or
three points, and the four-point invariants are non-nilpotent. The non-nilpotent invariants for any number of points, (or rather the elements of the quotient ring of invariants modulo the nilpotent ones), were constructed in \cite{hw3}. They have the property of being power series in $\l\p$. The nilpotent invariants start at five points and in $N=2$ they again have to be power series in $\l\p$ due to the $U(2)$ R-symmetry group. In $N=4$, however, this is not the case, because the R-symmetry group is $SU(4)$ not $U(4)$; the powers of $\l$ and $\p$ in any term in any invariant need only be the same modulo 4, the latter restriction being due to invariance under the $\bbZ_4$ centre of $SU(4)$. The existence of an $N=4$ five-point nilpotent invariant with leading term of the form $\l^4$ was demonstrated in \cite{ehw}.

In $N=4$ these results imply that the 4-point invariants, and indeed all of the non-nilpotent invariants, have an additional symmetry, $U_Y(1)$, which acts on the odd coordinates by $\d\l=\l,\ \d\p=-\p$ with the even coordinates being unchanged. Since this is also a symmetry of the basic propagators, it follows that $N=4$ $n$-point functions for $n<5$, are actually invariant under $PGL(4|4)$ and not just $PSL(4|4)$, as conjectured in \cite{ken1} and confirmed in \cite{ehw}.

The non-nilpotent invariants are much easier to construct explicitly, as they can be expressed in terms of superdeterminants and supertraces of coordinate difference matrices. For example, there are three independent invariants at four points in $N=2$ and they can be chosen to be

\be
S={\sdet X_{14}\sdet X_{23}\over\sdet X_{12}\sdet X_{34}} \; ,
\qquad T={\sdet X_{13}\sdet X_{24}\over\sdet X_{12}\sdet X_{34}}
\ee

and

\be
U=\str(X_{12}^{-1}X_{23} X_{34}^{-1} X_{41}) \; . \ee

They are all singular in the $y$ variables; for example, the leading terms in $S$ and $T$ are spacetime cross-ratios divided by internal cross-ratios.

%%%%%%%%%%%%%%%%%%%%%%%%%%%%%%%%%%%%%%%%%%%%%%%%%%%%%%%%%%%%%%%%%%%%

\section{2 and 3-point functions, anomalies and \newline
non-renormalisation}

The forms of the two- and three-point functions are determined
completely by superconformal invariance as one might expect. One has \cite{hw2}

\be
<p_1 p_2>\sim \d_{p_1 p_2}(g_{12})^{p_1} \ee

and

\be
<p_1 p_2 p_3>\sim (g_{12})^{p_{12}} (g_{23})^{p_{23}}
(g_{31})^{p_{31}} \ee

where

\be
p_{ij}=p_i+p_j-p_k \ee

and where $k\neq i,j$.

Let us now consider the field $\cO_2:=T$ in $N=4$. This field is
the $N=4$ supercurrent; it contains among its components the
traceless conserved energy-momentum tensor and the conserved
$SU(4)$ currents. These conservation laws and tracelessness
conditions arise because the superfield is constrained. On the
other hand, we know that even in a superconformal field theory some
at least of these conditions will be spoiled in the quantum theory
by anomalies, so that the three-point function cannot be analytic.
The resolution to this is that it is only formally analytic -
closer inspection reveals that it is necessary to regulate the $x$
singularities, and this is where the anomaly creeps in. For
example, we can extract from the above formula for the three-point
function the correlator for three $SU(4)$ currents. It agrees with
the expression obtained in components in \cite{dzf1}. Directly testing
the conservation of any one of these currents then produces a
non-zero answer due to the well-known triangle graph \cite{dzf1}.

In fact, the one-loop result for two and three-point functions of the supercurrent is exact as was shown in \cite{dzf1} using a result proven in \cite{afgj}. Another way of thinking about
this is to couple the theory to $N=4$ conformal supergravity \cite{hlt,hsw}.
Superspace non-renormalisation theorems indicate that the only
divergences in this theory occur at one loop \cite{hst,shelter}. Since these
divergences are directly related to the conformal anomaly \cite{duff} one
arrives at the same conclusion \cite{hsw}. This means that the coefficient of
the two- and three-point functions of $T$ cannot depend on the coupling - in other
words, the only contribution is at lowest order and all other
contributions vanish. 
%%%%%%%%%%%%%%%%%%%%%%%%%%%%%%%%%%%%%%%%%%%%%%%%%%%%%%%%%%%%%%%%%%%%

\section{The reduction formula and non-renormalisation for 2 and
3-point functions with arbitrary charges}

Although the anomaly argument given above shows that the
non-renormalisation theorem holds for the supercurrent, it has long
been suspected that this must be the case for arbitrary analytic
operators \cite{3pt}. One way of seeing this in the $N=4$ context is to use
the reduction formula which relates the derivative of an $n$-point
function to an $(n+1)$-point function with an insertion of the
integrated on-shell action. This formula was first proposed in the
current context in \cite{ken1}. Now the integrated on-shell action can
be written as a superaction in $N=4$ super Minkowski space \cite{hst2}, and
this translates into a certain integral in analytic superspace, so
that the reduction formula is

\be
{\del\over\del\t}<p_1\ldots p_n>\sim \int d\m_0 <T(X_0)p_1\ldots
p_n> \la{reduction} 
\ee

The measure $d\m$ is

\be
d\m:=d^4x d^4 y d^4\l \ee

or, in harmonic superspace notation, $d\m=d^4x du (D_{\a r'})^4$.

We have seen previously that two-, three- and four-point functions do not involve nilpotent invariants and therefore only involve the odd coordinates in the combination $\l\p$ (since the same is true for the propagators). For $n\leq 3$ any correlator on the left of \eq{reduction} has a non-vanishing leading term in a power series expansion in the odd variables and so in order for there to be a non-vanishing contribution to its derivative with respect to the coupling there would have to be a contribution to the integrand on the right of \eq{reduction} of the form $\l^4$ times some function of the even coordinates. Since there are no such terms we conclude that this derivative must vanish and hence that all two- and three-point analytic correlators are non-renormalised.

There is a slight loophole in the above argument and that is that
the integral on the RHS of \eq{reduction} means that contact terms could in
principle contribute \cite{petske}. Such local terms could either be associated
with the anomalies or they could be superconformally invariant
contact terms. In \cite{hssw} a search for invariant contact terms was made and none were found with the right fermion structure to contribute to the integral. We thus conclude that invariant contact terms cannot spoil the argument. On the other hand,
the anomalous terms, at least in the case of the energy-momentum
tensor, are already present in the three-point functions as we have seen.
Although one has to be careful when dealing with the contact points
arising from such expressions, it is nevertheless the case that
their fermion structure is such that they cannot contribute to 
this sort of integral. It seems likely that this situation will
also hold true for the higher-charge operators: any local terms
contributing on the RHS would have to have a different fermion
structure to that of the $U(1)_Y$-invariant expressions which are
valid for three or four separated points. There is no evidence that
invariant contact terms like this exist, and, on the other hand, it
is to be expected that the anomalies should arise from the formally
analytic expressions for separated points.

%%%%%%%%%%%%%%%%%%%%%EXTREMAL CORRELATORS%%%%%%%%%%%%%%%%%%%%%%%%%%

\section{Extremal correlators}

For four points or more the restrictions imposed on correlators by
analyticity are most evident when the charges of the operators are
chosen in an appropriate fashion. It was observed in AdS
supergravity that certain correlators seemed to be particularly
simple in such cases \cite{dfmnrext,arufrov,biakov}. A test of the Maldacena conjecture, therefore, is to check whether this is the case on the field theory side. These so-called extremal correlators are those
for which one of the charges $p_1$, say, is equal to the sum of all
of the others,

\be
 p_1=\sum_{i=2}^n p_i
\ee

In the SCFT the prefactor $P$ in $<p_1\ldots p_n>=P F$ can be
written uniquely as

\be
 P=\prod_{i=2}^n (g_{1i})^{p_i}
\ee

This in turn means that the function of invariants which multiplies
the prefactor is only allowed to have singularities of a rather
special type. Typically, the singularities that arise in the
invariants are of the form $(y_{12}y_{34})^{-1}$ whereas the zeroes
in $P$ involve products of $y$'s where each factor has a subscript
1. This rather rough argument indicates how it can be that the only
functions of invariants allowed for in extremal correlators are in
fact constant. It is confirmed in more detail in \cite{ehssw} for four points and extended to $n$ points in \cite{ehsw} using $N=2$ superfieldsand the
reduction formula which also allows one to show that the constant
does not depend on the coupling, in a similar fashion to the
non-renormalisation theorems for two- and three-point functions.
Moreover, it is also possible to show that next-to-extremal
correlators, for which $p_1=(\sum_{i=2}^n p_i) -2$, are also
trivial and non-renormalised, and this has been confirmed on the supergravity side \cite{erdPer}. Similar results have been shown in M-theory on $AdS_{7(4)} \xz S^{4(7)}$ \cite{dhp} and seem also to be true on the field theory side in
$D=6 (2,0)$ SCFT \cite{hd6}.

%%%%%%%%%%%%%%%%%%%%%%4PTS%%%%%%%%%%%%%%%%%%%%%%%%%%%%%%%%%%%%%%%%

\section{4-point functions}

Initially the analytic superspace programme focused on the analysis
of correlators of four charge two hypermultiplet composites in
$N=2$, and it was conjectured that for such low charges it might be
possible to determine such functions completely by symmetry and
analyticity. However, this has turned out not to be the case. We
shall not give the details here but outline the main points.

The correlator to be studied is $<2222>$, although the operators
need not be identical (there are three such charge two operators in
$N=4$, for example). It can be written in the form

\be
<2222>=(g_{12})^2 (g_{34})^2 F \ee

where $F$ is a function of invariants. At four points in $N=2$
there are  three independent super-invariants which we discussed above and which correspond to the three purely bosonic invariants which arise at
lowest order. These are two $x$ cross-ratios $s,t$ and one $y$
cross-ratio $v$, where

\be
s={x_{14}^2 x_{23}^2\over x_{12}^2 x_{34}^2} \; , \qquad
t={x_{13}^2 x_{24}^2\over x_{12}^2 x_{34}^2} \ee

and

\be
v={y_{14}y_{23}\over y_{12} y_{34}} \ee

Due to the fact that the propagators behave as
$(y_{12})^2(y_{34})^2$ it follows that \cite{ehpsw}

\be
F=a_1(S',T') + a_2(S',T')V + a_3(S',T') V^2 \ee

where $S',T'$ are $V$ are the three functions which (uniquely)
extend $s,t$ and $v$ to super-invariants. Explicitly,

\be
S'=SV,\qquad T'=T(1+V),\qquad V={T+U-1\over 1+S-T} \ee

where the simpler invariants $S,T$ and $U$ are given in equations (24) and (25).

The idea is to expand $F$ in a power series in odd variables and
study each order to see what restrictions, if any, are placed on
the functions $a_i$. This can be done using a set of partially
supersymmetric variables $\L,\P$ which are also analytic in $y$. In
terms of these variables the expansion goes up to fourth order in
$\L\P$, but the only constraints that arise occur at first order.
Alternatively, one can use harmonic variables. One finds that two
combinations of the three functions $a_i$ satisfy two coupled
partial differential equations while $a_3$ is unconstrained \cite{ehpsw}.

Although this result is not as strong as originally had been hoped,
nevertheless analyticity does yield results at first order which go
beyond what might expect from naive symmetry considerations.
Recently this result has been slightly strengthened, in the $N=4$ case, using symmetry considerations (under interchange of points) and the reduction formula \cite{epss}. The result is that the entire amplitude of four supercurrents is determined in terms of one function of $s$ and $t$.

%%%%%%%%%%%%%%%%%%%%%%%%%%%%%%OPE%%%%%%%%%%%%%%%%%%%%%%%%%%%%%%%%%%

\section{The OPE and analytic superspace}

In \cite{hw4} the OPE of $N=4$ analytic operators was investigated
and it was claimed that this should close in the sense that only such operators and their derivatives
appear on the right-hand side. As we shall see, this is indeed the case, provided that we consider more general analytic tensor superfields, although at the  time it was not clear how operators such as the Konishi operator could be accommodated in this formalism. In \cite{afgj2} it was shown that this operator occurs in the OPE of two supercurrents of any $N=1$ SCFT. Now the $N=4$
Konishi multiplet is $\tr(W_{ij}W^{*ij})$ and does not appear to be
analytic; in fact, it is a long multiplet. However, it is very easy
to see using harmonic superspace notation that it does occur in the
OPE of two $N=4$ supercurrents even in the free theory. Consider
then the free theory where $T=W^2$ and $W=u^{ij}W_{ij}$, with
$u^{ij}:=\e^{rs}u_r{}^i u_s{}^j$. $K$ is absent in the product $W^2$ due
to the fact that it can be written as ${1\over2}\e^{ijkl}W_{ij}W_{kl}$, and
so drops out because of the $u$'s. Now consider the OPE of two
$T$'s; it will include a contribution which involves a single
contraction, and this for the free theory is simply the propagator
$g_{12}$. Hence we find

\be
T(1)T(2)\sim g_{12} W(1) W(2) + \ldots \ee

and

\bea
 W(1) W(2)&=&u(1)^{ij} u(2)^{kl} W_{ij}(1)W_{kl}(2) \nn\\
 &=&u(1)^{ij} u(2)^{kl} \left(W_{ij}(1)W_{kl}(1) + \ldots\right)\nn \\
 &= &u(1)^{ij} u(2)^{kl}\left( T_{ij,kl}(1) + \e_{ijkl} K(1) + \ldots\right)\nn
\eea

But, since the $u$'s are at different points we can no longer
conclude that they are annihilated when contracted with
$\e_{ijkl}$. Thus the Konishi operator is certainly present in the
$N=4$ OPE. At first sight, this seems to be a disaster for the
analytic superspace method because the OPE seems to produce
operators which are no longer analytic. However, this is not the
case. It turns out that the Konishi operator can be written as an
operator on analytic superspace provided that one is prepared to
include more general superfields. It is simply an example of a
superfield which has superindices, i.e. an analytic tensor
superfield.

To see this explicitly consider the OPE for two $T$'s in analytic
superspace. Looking again at single contractions we shall have a
term

\be
T(1)T(2)\sim g_{12} W(1) W(2) \ee

but now we regard the $W$'s as functions of the analytic
coordinates. We can now make a Taylor expansion about point $2$ in
all the coordinates. Up to second order this gives

\be
W(1)W(2)=T + X_{12}^{AA'}(\del_{AA'}W) W + {1\over2}
X_{12}^{BB'}X_{12}^{AA'}(\del_{AA'}\del_{BB'}W) W +\ldots \ee

The second term can clearly be arranged to give $1/2 X_{12}\cdot
\del T$, but there are obviously going to be new operators at
second and higher orders which have indices. By taking linear
combinations of operators with the same number of derivatives one
can combine all the operators with a given number of derivatives
into derivatives of known operators together with operators which
are quasi-primary, i.e that transform in a tensorial fashion under
the superconformal group.

One of the operators that arises in this way at second order is 

\be
\cO_{AB,A'B'}=\del_{(A(A'}W \del_{B)B')}W-{1\over6}\del_{(A(A'}\del_{B)B')}(W^2)
\la{kon} \ee

where the round brackets denote generalised symmetry on the primed and unprimed indices. This means
that it is symmetric on the spacetime spinor indices and antisymmetric on the
internal indices. The lowest dimensional component of this operator
is proportional to

\be
\e^{ab}\e^{a'b'}\left(\del_{aa'}W\del_{bb'}W-{1\over6}
\del_{aa'}\del_{bb'}(W^2)\right) \sim K \ee

So the operator \eq{kon} is simply the Konishi operator written as an
analytic tensor field. Indeed, it can be shown that all
superconformal representations (at least of the type that are of
interest in field theory) can be represented on analytic superfields \cite{hesh2}.

The conclusion, therefore, is that the analytic OPE does contain more terms than
was suggested in \cite{hw4} and does indeed contain multiplets such as the Konishi multiplet on the
right-hand side. On the other hand, it is not incompatible with
analyticity, at least in the free theory. The situation is more complicated in the interacting case but this simple example gives us hope, even in this case,  that the OPE and analyticity will remain compatible. It would be interesting to pursue this topic further, particularly in the light of recent results on the $N=4$ OPE and the Konishi multiplet \cite{afp,aeps,bkrs}.

%%%%%%%%%%%%%%%%%%%%%%%%%%%%%%%%%%%%%%%%%%%%%%%%%%%%%%%%%%%%%%%%%%%

\section{Conclusions}

To summarise we have seen that superspace methods give a very explicit and clear way of implementing the group-theoretical aspects of the Maldacena conjecture for $AdS_5\xz S^5$. On the field theory side, harmonic superspace methods provide the easiest way of implementing the superconformal Ward identities. These identities, together with analyticity, enable us to prove, almost rigorously, the non-renormalisation theorems for two- and three-point functions, and to establish the triviality of extremal and next-to-extremal correlations functions. Further simplifications in other correlators are also implied. Finally, the OPE seems to be consistent with analyticity, since apparently non-analytic operators such as the Konishi operator can be interpreted as more complicated analytic superfields.

%%%%%%%%%%%%%%%%%%%%%%%%%%%%%%%%%%%%%%%%%%%%%%%%%%%%%%%%%%%%%%%%%%

\section*{Acknowledgements}

This work was supported in part by PPARC through SPG grant 68.

We would like to thank our collaborators B. Eden, C. Schubert, E.
Sokatchev.

We would also like to thank the following for comments, discussions
and critical remarks: M. Bianchi, E. d'Hoker, D.Z. Freedman, M.
Grisaru, K. Intriligator, K. Skenderis.

\end{document}